\documentstyle[12pt]{article} 
 
\def\kbar{\overline{k}} 
\def\dbar{\overline{\partial}} 
 
\def\alphabar{\overline{\alpha}} 
\def\zbar{\overline{z}} 
\def\abar{\overline{\alpha}} 
\def\bbar{\overline{\beta}} 
\def\ghat{\hat{g}} 
\def\delbar{\overline{\partial}} 
\def\kbar{\overline{k}} 
\def\gambar{\overline{\gamma}}

\begin{document}

\begin{titlepage} 
 
\begin{flushright} 
QMW-PH-97-31\\ 
hep-th/9710096\\
revised 
\end{flushright} 
 
\vspace{3cm} 
 
\begin{center} 
 
{\bf \Large Conformal Invariance and Duality in Self-Dual Gravity and (2,1) 
Heterotic String Theory}

\vspace{.7cm} 
 
M.\ Abou Zeid and C.\ M.\ Hull 
 
\vspace{.7cm} 
 
{\em Physics Department, Queen Mary and Westfield College, \\ 
Mile End Road, London E1 4NS, U.\ K.\ } 
 
\vspace{3cm} 
 
October 1997 
 
\vspace{1cm}

\begin{abstract} 
 
A system of gravity coupled to a 2-form gauge field, a dilaton and Yang-Mills 
fields in $2n$ dimensions arises from the (2,1) sigma model or string. The 
field equations imply that the curvature with torsion and Yang-Mills field 
strength are self-dual in four dimensions, or satisfy generalised self-duality 
equations in $2n$ dimensions. The Born-Infeld-type action describing this 
system is simplified using an auxiliary metric and shown to be 
classically Weyl invariant only in four dimensions. A dual form of the action 
is found (no isometries are required). In four dimensions, the dual geometry is 
self-dual gravity  without torsion coupled to a scalar field. In $D>4$ 
dimensions, the dual geometry is hermitian and determined by a 
$D-4$ form potential $K$, generalising the K\"{a}hler potential of the four 
dimensional case, 
with the fundamental 2-form given by $\tilde J= i*\partial \bar \partial K$. The coupling to
Yang-Mills is through a term $K\wedge tr (F\wedge F)$ and leads to a Uhlenbeck-Yau
field equation $\tilde J^{ij}F_{ij}=0$.
 
\end{abstract} 
 
\end{center} 
 
\end{titlepage}

\section{Introduction}

        The superstring with (2,1) world-sheet supersymmetry provides important 
insights into M-theory and 
superstring theory. The target space of the (2,1) string is 2+2 dimensional, 
with a null reduction restricting the dynamics to 1+1 or 2+1 
dimensions~\cite{OV}. 
The target space dynamics has been shown in~\cite{KM1,KM2,M1} to describe 
critical string worldsheets or membrane worldvolumes in static gauge and 
constitutes an explicit realisation of the scenario proposed in~\cite{MBG}. 
Furthermore, it was found that all types of ten-dimensional superstring 
theories 
and the eleven-dimensional supermembrane arise as vacua of the (2,1) heterotic 
string. More recently, 
Martinec~\cite{M2,M3} has argued that the (2,1) string may provide the degrees of 
freedom needed to 
define certain compactifications of the matrix model of M-theory proposed in 
ref.~\cite{BFSS}. 
 
        The (2,1) heterotic string was shown in~\cite{OV} to describe a 
theory of gravity with torsion coupled to Yang-Mills gauge fields in 2+2 
dimensions. The null reduction mentioned above must be imposed, and 
yields a 1+1 dimensional space or a 2+1 dimensional space  
depending on the orientation of the 
null Killing vector used in  the null reduction~\cite{OV}. The field equations 
were found 
in~\cite{H2,H0} and the effective action for the gravitational and Yang-Mills 
degrees of freedom 
(before null reduction) was obtained   in refs.~\cite{KM3,H1}. The geometry 
is a generalisation of K\"{a}hler geometry with torsion~\cite{HW} and a 
hypersymplectic 
structure~\cite{H1}, and the field equations imply that the curvature with 
torsion is self-dual in 2+2 dimensions. The Yang-Mills fields are also 
self-dual 
in 2+2 dimensions. In higher dimensions, the field equations imply that the 
curvature with torsion has $SU(n_1 ,n_2 )$ holonomy, while the Yang-Mills 
fields satisfy a non-linear form of the Uhlenbeck-Yau equation~\cite{H1}.  
The action in 10+2 dimensions is the effective space-time theory that is 
conjectured to give supergravity in 10+1 or 9+1 dimensions upon null  
reduction~\cite{KM3}. 
 
        Our purpose here is twofold. In section~\ref{sec:pol}, we  
will formulate an equivalent form of the (2,1) string action with an auxiliary metric  
and will  show that  it 
is Weyl invariant only in four dimensions. In section~\ref{sec:dual}, we will 
dualise the vector potential that governs the geometry to find an equivalent 
action given in terms of a $D-4$ form potential in $D$ dimensions. In four 
dimensions, the dual geometry is K\"{a}hler.

\section{(2,1) Geometry}			\label{sec:geom}

        We begin by recalling the geometric conditions for (2,1) supersymmetry 
of the (1,1) sigma model with metric $g_{ij}$ and anti-symmetric 
tensor $b_{ij}$~\cite{HW,GHR} (see also refs.~\cite{H0,H2}; further discussion 
of the geometry, isometry symmetries and gauging of the (2,1) model can be 
found 
in refs.~\cite{AH1,AH2,AH3}). The sigma model 
is invariant under (2,1) supersymmetry~\cite{HW,H2,H0} if the target space is 
even dimensional ($D=2n$) with a complex structure $J^i{}_j$ which is 
covariantly constant with respect to the connection with torsion $\Gamma^{(+)}$ 
and with respect to which the metric is hermitian, so that $J_{ij} \equiv 
g_{ik}J^k{}_j$ is antisymmetric.

        It is useful to introduce complex coordinates $z^\alpha ,\zbar^{\bbar}$ 
in which the line element is $ds^2 = 2g_{\alpha \bbar} dz^\alpha 
d\zbar^{\bbar}$ and the exterior derivative decomposes as $d=\partial +\delbar 
$. The conditions for (2,1) supersymmetry imply that $H$ is given in terms of 
the fundamental 2-form 
\begin{equation} 
J=\frac{1}{2}J_{ij}d \phi^i \wedge d\phi^j = -ig_{\alpha \bbar} dz^\alpha \wedge 
d\zbar^{\bbar} 
\end{equation} 
by 
\begin{equation} 
H= i(\partial -\delbar )J . 
\end{equation} 
Then the condition $dH=0$ implies 
\begin{equation} 
i\partial \delbar J =0   \label{dj0} 
\end{equation} 
so that locally there is a (1,0) form potential $k=k_\alpha dz^\alpha$ such that 
\begin{equation} 
J=i(\partial \kbar +\delbar k ) . 
\label{21&J} 
\end{equation} 
In a suitable gauge, the metric and torsion potential are then given by 
\begin{equation} 
g_{ij} = \left( \begin{array}{cc} 0 & g_{\alpha \bbar} \\ g_{\alphabar \beta} 
& 0 \end{array} \right) \ \ , \ \ b_{ij} = \left( \begin{array}{cc} 0 & 
b_{\alpha 
\bbar} \\ b_{\alphabar \beta} & 0 \end{array} \right)  
\label{gstruct} 
\end{equation} 
so that the torsion (2) is given by 
\begin{equation} 
H_{\alpha \beta  \gambar} =\frac{1}{2} \left( g_{\alpha \gambar , \beta}  
-g_{\beta \gambar ,\alpha } \right) \ \ ,\ \ H_{\alpha \beta \gamma} = 0 
\end{equation} 
while the constraint (3) implies that the metric satisfies 
\begin{equation} 
g_{\alpha [ \bbar ,\gambar ] \delta} -g_{\delta [ \bbar ,\gambar ] \alpha } =0 . 
\label{2ndconstr} 
\end{equation} 
 
	From (4), the geometry is defined locally by 
\begin{eqnarray} 
g_{\alpha \bbar} & = & \partial_\alpha \kbar_{\bbar} +\delbar_{\bbar} k_\alpha 
\nonumber \\ b_{\alpha \bbar} & = & \partial_\alpha \kbar_{\bbar} - 
\delbar_{\bbar} k_\alpha . 
\label{21geom} 
\end{eqnarray} 
If $k_\alpha =\partial_\alpha K$ for some $K$, then the torsion vanishes and 
the manifold is K\"{a}hler with K\"{a}hler potential $K$, but if $dk \neq 0$ 
then the space is a hermitian manifold of the type introduced in~\cite{HW}.

        The (1,1) supersymmetric model will be conformally invariant at 
one-loop if there is a function $\Phi$ such that 
\begin{equation} 
R_{ij}^{(+)} -\nabla_{(i}\nabla_{j)} \Phi -H^k{}_{ij} \nabla_k \Phi =0 
\label{oloop} 
\end{equation} 
where $R_{ij}^{(+)}$ is the Ricci tensor for the connection with torsion. The 
curvature and Ricci tensors with torsion are 
\begin{equation} 
R^{(+)k}_{lij} = \partial_i \Gamma^{(+)k}_{jl} -\partial_j \Gamma^{(+)k}_{il} 
+\Gamma^{(+)k}_{im}\Gamma^{(+)m}_{jl} -\Gamma^{(+)k}_{jm}\Gamma^{(+)m}_{il}  
, 
\ \ R^{(+)}_{ij} = R^{(+)k}_{ikj} . 
\label{RimRic} 
\end{equation} 
 
        It will be useful to define the vector 
\begin{equation} 
v^i=H_{jkl}J^{ij}J^{kl} 
\end{equation} 
together with  the $U(1)$ part of the curvature 
\begin{equation} 
C^{(+)}_{ij}=J^l{}_kR^{{(+)} k} {}_{lij} 
\label{defC} 
\end{equation} 
and the $U(1)$ part of the connection 
\begin{equation} 
\Gamma^{(+)} _i=J^k {}_j\Gamma^{(+)j}_{ik}=i( \Gamma^{(+)\alpha}_{i \alpha}- 
\Gamma^{+\bar \alpha}_{i\bar \alpha}) . 
\end{equation} 
In a complex coordinate system, (\ref{defC}) can be written as 
\begin{equation} 
C^{(+)} _{ij}= \partial _i \Gamma^{(+)} _j - \partial_j \Gamma^{(+)} _i . 
\end{equation} 
If the metric has Euclidean signature, then the holonomy of any metric 
connection (including $\Gamma^{(+)}$) 
is contained in $O(2n)$, while if it has signature $(2n_1,  2n_2)$ where 
$n_1+n_2=n$, it will be in 
$O(2n_1,2n_2)$. The holonomy ${\cal H}(\Gamma^{(+)})$ of the connection 
$\Gamma^{(+)}$ is contained 
in $U(n_1,n_2)$. It will be contained in $SU(n_1,n_2)$ if in addition 
\begin{equation} 
C^{(+)} _{ij}=0 
\label{C=0} 
\end{equation} 
where the $U(1)$ part of the curvature is given by~(\ref{defC}). 
As $C_{ij}$ is a representative of the first Chern class, a necessary condition 
for 
this is the 
vanishing of the first Chern class. 
 
It was shown in~\cite{H2} that geometries for which 
\begin{equation} 
\Gamma_{i}^{(+)} =0 \label{Gam0} 
\end{equation} 
in some suitable choice of coordinate system will satisfy the one-loop 
conditions~(\ref{oloop}) provided the dilaton is chosen as 
\begin{equation} 
\Phi = -\frac{1}{2} \log | \det g_{\alpha \bbar} | , 
\end{equation} 
which implies 
\begin{equation} 
\partial _i \Phi= v_i . 
\end{equation} 
 Moreover, the one-loop dilaton 
field equation is then satisfied for compact manifolds, or for non-compact ones 
in which $\nabla 
\Phi$ falls off sufficiently fast~\cite{H1}. This implies that ${\cal 
H}(\Gamma^{(+)}) 
\subseteq SU(n_1,n_2)$ and these geometries  
generalise the K\"{a}hler Ricci-flat or 
Calabi-Yau 
geometries, and reduce to these in the special case in which 
$H=0$. 
These are not the most general solutions of (\ref{oloop})~\cite{H1}. 
 
The condition that the connection $\Gamma^{(+)}$ has $SU(n_1,n_2)$ holonomy 
can be cast as a 
generalised self-duality condition on the curvature $R^{(+)}$. 
Defining the four-form 
\begin{equation} 
\phi^{ijkl} \equiv -3 J^{[ij}J^{kl]} 
\label{star} 
\end{equation} 
the condition that ${\cal 
H}(\Gamma^{(+)}) 
\subseteq SU(n_1,n_2)$ 
is equivalent to 
\begin{equation} 
R^{(+)}_{ijkl} = \frac{1}{2}g_{im}g_{jn}\phi^{mnpq}R^{(+)}_{pqkl} . 
\label{sdR} 
\end{equation} 
For $D=4$, $\phi^{ijkl}= -\epsilon ^{ijkl}$ and this is the usual 
anti-self-duality condition, 
while for $D>4$, this 
 is an example of the  generalised  self-duality equations considered for 
Riemannian 
manifolds   in refs.~\cite{Bon,CDFN,DGT,AL}.

        The equation~(\ref{Gam0}) can be viewed as a field equation for the 
potential 
$k_\alpha$. It can be obtained by varying the action~\cite{KM3,H1} 
\begin{equation} 
S=\int d^D x \sqrt{ | \det g_{\alpha \bbar } | }     \label{KMHact} 
\end{equation} 
where $g_{\alpha \bbar}$ is given in terms of $k_\alpha$ by~(\ref{21geom}). 
It follows from the form~(\ref{gstruct}) of the metric that this action can be 
rewritten as 
\begin{equation} 
S= \int d^D x | \det g_{ij} |^{1/4}       \label{KMH4} 
\end{equation} 
which is non-covariant but is invariant under volume-preserving 
diffeomorphisms.

This can be generalised to include Yang-Mills fields $A_i$ taking values in 
some group $G$, 
in addition to $g_{ij}$ and $b_{ij}$. 
The  $A_i$ must be a connection for a holomorphic vector bundle (so that the 
field-strength $F$ 
is a (1,1) form), with Chern-Simons form $\Omega (A)$ satisfying $d\Omega =tr 
F^2 $, 
Bott-Chern form $\Upsilon$~\cite{BC} defined by 
\begin{equation} 
tr (F^2 ) =i\partial \delbar \Upsilon  
\label{F2&U}
\end{equation} 
and a form $\chi$ defined by 
\begin{equation} 
\Omega (A) = i(\partial -\delbar )\Upsilon +d\chi . 
\end{equation} 
The conditions for (2,1) 
supersymmetry and conformal invariance then imply the existence of a (1,0) form 
$k$ such that~(\ref{21&J}) is replaced by 
\begin{equation} 
J = \Upsilon  +i ( \partial \kbar +\delbar k ) 
\label{21&JYM} 
\end{equation} 
and the metric and torsion potential are given by 
\begin{eqnarray} 
g_{\alpha \bbar} & = & i\Upsilon_{\alpha \bbar} + \partial_{\alpha} 
\kbar_{\bbar} +\dbar_{\bbar} k_{\alpha} \nonumber \\ b_{\alpha \bbar} & = & 
i\chi_{\alpha \bbar} +\partial_{\alpha} k_{\bbar} 
-\dbar_{\bbar}  k_{\alpha} . 
\label{mod21geom} 
\end{eqnarray} 
The field equations can again be obtained by varying the 
action~(\ref{KMHact}), but with $g_{\alpha \bbar}$ given by~(\ref{mod21geom}). 
The Yang-Mills equation is 
\begin{equation} 
J^{ij}F_{ij} =0 . 
\end{equation}

        It is sometimes useful to write the metric in terms of a fixed 
background metric 
$\hat{g}_{\alpha \beta}$ (e.~g.\ a flat metric) which is given in terms of a 
potential $\hat{k}$ by $\hat{g}_{\alpha \bbar} =  
\delbar_{\bbar} \hat{k}_{\alpha} 
+\partial_{\alpha} 
\hat{\kbar}_{\bbar}+i\Upsilon_{\alpha \bbar}$, and a fluctuation given 
in terms of a vector field $B_i$ defined by 
\begin{equation} 
B_{\alpha} = i(k_\alpha -\hat{k}_{\alpha}) \ \ ,\ \ B_{\abar} =-i(\kbar_{\abar} 
-\hat{\kbar}_{\abar})  
\label{defB} 
\end{equation} 
with field strength ${\cal F}_{ij}=2\partial_{[i}B_{j]}$. Then 
\begin{equation} 
g_{\alpha \bbar}  =  \hat{g}_{\alpha \bbar} 
+i{\cal F}_{\alpha \bbar} . 
\label{defh} 
\end{equation} 
The  
action~(\ref{KMHact}) becomes 
\begin{equation} 
S= \int d^D x \sqrt{ | \det (\hat{g}_{\alpha \bbar} + i{\cal F}_{\alpha \bbar}) 
| } 
\label{KMHact'} 
\end{equation} 
which is similar to a Born-Infeld action 
and is invariant under the abelian  gauge symmetry 
$\delta B=d\lambda$.

\section{Weyl-Invariant Action for (2,1) Strings}		\label{sec:pol}

The Nambu-Goto action for the bosonic string can be rewritten using an auxiliary world-sheet metric 
~\cite{BVH,HT} in a way which is useful for many purposes, such as quantization~\cite{P}. 
	Similarly, the action~(\ref{KMH4}) can be written in the classically  
equivalent alternative form 
\begin{equation} 
S' = T'_4  \int d^D x |\gamma|^{1/4} \left[ 
\gamma^{ij}  g_{ij}    -(D-4)c \right]  \label{Hform4} 
\end{equation} 
where $\gamma_{ij}$ is an auxiliary metric, $\gamma =\det  
\gamma_{ij}$ and $c,T'$ are (real)  
constants. The field equation for $\gamma_{ij}$ is 
\begin{equation} 
\gamma_{ij} =\frac{1}{c}  g_{ij}   
\label{fe} 
\end{equation} 
for $D\neq 4$, and  
\begin{equation} 
\gamma_{ij} = \frac{4}{(\gamma^{kl}g_{kl})}g_{ij} 
\label{fe4} 
\end{equation} 
for $D=4$. Substituting back in~(\ref{Hform4}) one recovers the  
action~(\ref{KMH4})  
with the constant $T'$ given by 
\begin{equation} 
T'_4 = \frac{1}{4} c^{\frac{D}{4}-1} . 
\end{equation} 
 
        In complex coordinates, the action~(\ref{Hform4}) takes the form 
\begin{equation} 
S' = T' \int d^D x \sqrt{| \gamma |} \left[ \gamma^{\alpha \bbar} 
g_{\alpha \bbar} +\gamma^{\alphabar \beta} 
g_{\alphabar \beta}  -(D-4)c \right] 
\label{Hform} 
\end{equation} 
where now $\gamma = \det \gamma_{\alpha \bbar}$.  
The field equations~(\ref{fe}) or~(\ref{fe4}) imply that the  components $\gamma^{\alpha \beta}$ and 
$\gamma^{\abar \bbar}$ vanish, so that on-shell 
the auxiliary  
metric $\gamma_{ij}$ is hermitian, 
\begin{equation} 
J^j{}_{(i}\gamma_{ k) j} =0 . 
\label{constgamm} 
\end{equation} 
It is then consistent to impose the condition~(\ref{constgamm}) that the metric be hermitian off-shell as 
well, and we shall do so in what follows. 
 
	The action~(\ref{Hform})  is a special case  
of the general class of action 
\begin{equation} 
S' = T'_q \int d^D x |\gamma|^{1/q} \left[ \gamma^{ij}  g_{ij} 
 -(D-q)c \right] .    \label{Hgen} 
\end{equation} 
For $D\neq q$, the field equation for the auxiliary tensor  
is~(\ref{fe}), and substituting this back in~(\ref{Hgen})  
yields actions of the form 
\begin{equation} 
S = \int d^D x | \det (g_{ij} ) |^{1/q}         \label{KMH4'} 
\end{equation} 
with the constant $T'$ given by 
\begin{equation} 
T'_q = \frac{1}{q} c^{\frac{D}{q}-1} . 
\end{equation} 
 
	In the special case in which $D=q$, the constant  
term in the action~(\ref{Hgen})  
vanishes and there is a generalised Weyl symmetry under 
\begin{equation} 
\gamma_{ij} \rightarrow \omega (x) \gamma_{ij} .        \label{Weyl} 
\end{equation} 
The field equation in this case is 
\begin{equation} 
\gamma_{ij} = \frac{q}{(\gamma^{kl}g_{kl})}g_{ij} . 
\end{equation} 
 
	For all $D,q$ there is in addition an  
invariance under volume preserving 
diffeomorphisms, 
i.e. diffeomorphisms of the $D$-dimensional space-time which preserve  
$\det \gamma_{ij}$,  
so that the vector field $\xi^i$  generating 
the  diffeomorphism 
must satisfy 
$\nabla _i\xi^i=0$ 
where $\nabla _i\xi^i= \gamma ^{-1/2}\partial _i(\gamma ^{1/2} \xi^i)$ and 
$\nabla_i$ is the usual 
covariant derivative 
 for the metric  $\gamma_{ij}$. 
For $D=q$, the symmetry consists of the volume preserving 
diffeomorphisms, 
 together with the Weyl transformations.

\section{Duality}				\label{sec:dual}

        We now discuss the dualisation of the vector potential  $k_i$, 
starting with the simplest case of four space-time dimensions and no  
background metric or Yang-Mills fields. The discussion 
will be generalised below to include the background metric and the coupling to 
the Yang-Mills fields. 
 
	Consider the action 
(\ref{KMHact}) and 
add a Lagrange multiplier term imposing the constraint~(\ref{21geom}), 
\begin{equation} 
S = \int d^4 x \left[ \sqrt{ |g| } 
-\frac{1}{4} 
\left\{ 
 \Lambda^{\alpha \bbar} \left( g_{\alpha \bbar} -\partial_\alpha 
\kbar_{\bbar} 
-\delbar_{\bbar} k_\alpha \right) +c.c. \right\} 
\right] 
\label{act1} 
\end{equation} 
with $g \equiv \det  
g_{\alpha \bbar}$ (the sign of the Lagrange term is arbitrary). Eliminating 
$\Lambda^{\alpha \bbar}$ from~(\ref{act1}), we recover the  
action~(\ref{KMHact}) subject to the constraint~(\ref{21geom}).  
Alternatively, we can first integrate over the  
vectors  
$k_{\alpha}$, $\kbar_{\alphabar}$, which are   Lagrange multipliers for the  
constraints 
\begin{eqnarray} 
\partial_\alpha \Lambda^{\alpha \bbar} & = & 0 \nonumber \\ \delbar_{\bbar} 
\Lambda^{\alpha \bbar} & = & 0 . 
\label{constr} 
\end{eqnarray} 
In $D=4$ dimensions, these can be solved locally in terms of a scalar $K$~: 
\begin{equation} 
\Lambda^{\alpha \bbar} = L^{\alpha \bbar} 
\label{solconstr} 
\end{equation} 
where $L^{\alpha \bbar}$ is the \lq field strength' of $K$  given by
\begin{equation} 
L^{\alpha \bbar} \equiv \epsilon^{\alpha \gamma \bbar \overline{\delta}} 
\partial_\gamma \delbar_{\overline{\delta}} K  \label{defL} 
\end{equation} 
and $\epsilon^{\alpha \gamma \bbar \overline{\delta}}$ is the 
antisymmetric tensor density (with $\epsilon^{1 \overline{1} 2  \overline{2 }}=1$). Then integrating over 
$k,\kbar$ and 
solving as in~(\ref{solconstr}), the action takes the form 
\begin{equation} 
S= \int d^4 x \left[ \sqrt{|g|} -\frac{1}{4} \left\{  
L^{\alpha \bbar} g_{\alpha 
\bbar} 
+c.c. \right\} \right] . 
\label{actg+L} 
\end{equation} 
Note that in (\ref{act1}) we have chosen $\Lambda^{\alpha \bbar}$ to be a tensor density so that the 
second term in the action is fully diffeomorphism invariant, even though the first term is only invariant 
under volume preserving diffeomorphisms. This proves to be the most convenient choice, but equivalent 
results could have been obtained by choosing $\Lambda^{\alpha \bbar}$ to transform differently, so that 
the second term in 
(\ref{act1}) was also only invariant under volume preserving diffeomorphisms. 
 
	Alternatively, one can add a Lagrange multiplier term imposing  
the constraint~(\ref{2ndconstr}), 
\begin{equation} 
S = \int d^4 x \left[ \sqrt{| g| } 
-\frac{1}{4} \left\{ 
\epsilon^{\alpha \gamma \bbar \overline{\delta}} K  
\partial_\gamma \partial_{\bar \delta}  
g_{\alpha \bbar} 
+c.c. \right\} \right] \label{Salt}
\end{equation} 
Integrating by parts yields 
\begin{equation} 
S = \int d^4 x \left[ \sqrt{| g| } 
-\frac{1}{4}\left\{ 
 \epsilon^{\alpha \gamma \bbar \overline{\delta}}  
\partial_\gamma \partial_{\bar \delta} K g_{\alpha \bbar} 
+c.c. \right\} 
 \right] 
\end{equation} 
which by definition~(\ref{defL}) of $L^{\alpha \bbar}$ is identical  
to action~(\ref{actg+L}). Thus it is equivalent 
to impose either of the constraints~(\ref{21geom}) or~(\ref{2ndconstr}). 
   
The field equation for $g_{\alpha \bbar}$ which follows from~(\ref{actg+L}) is 
\begin{equation} 
\sqrt{|g|} g^{\alpha \bbar} = L^{\alpha \bbar} . 
\label{fegonly} 
\end{equation} 
Taking determinants in~(\ref{fegonly}) yields the constraint 
\begin{equation} 
\det L^{\alpha \bbar} =  -1   \label{constraint} 
\end{equation} 
for signature (2,2) or $\det L^{\alpha \bbar} = 1$ for signature (4,0). 
 Taking the trace gives 
\begin{equation} 
L^{\alpha \bbar} g_{\alpha \bbar} = 2\sqrt{|g|} . 
\label{traceg} 
\end{equation} Substituting~(\ref{traceg}) back into the action~(\ref{actg+L}), 
a cancellation 
occurs and the action vanishes, 
\begin{equation} S=0 . 
\end{equation} The dynamics is contained entirely in the 
constraint~(\ref{constraint}). Consider the K\"{a}hler metric $G_{\alpha  
\bbar}$ with potential $K$, 
\begin{equation} 
G_{\alpha \bbar} \equiv  \partial_\alpha \delbar_{\bbar} K . 
\label{defG} 
\end{equation} 
Then~(\ref{constraint}) implies 
\begin{equation} 
\det G_{\alpha \bbar} =- 1  
\label{constrG} 
\end{equation} 
for signature (2,2), or $\det G_{\alpha \bbar} = 1 $ for signature (4,0). 
Thus the dual metric $G_{\alpha \bbar}$ is K\"{a}hler and Ricci-flat.  

The equation~(\ref{fegonly}) implies
\begin{equation}
g_{\alpha \bbar} = \Omega L_{\alpha \bbar}
\end{equation}
for some scalar field $\Omega$, and the constraint~(\ref{2ndconstr}) will be satisfied if~(\ref{constrG}) and
\begin{equation}
G^{\alpha \bbar} \partial_\alpha \partial_{\bbar} \Omega = 0
\label{feO}
\end{equation}
hold. Then $\Omega$ is a harmonic scalar on the dual space.

Writing  
$G_{\alpha \bbar}= \eta_{\alpha \bbar}+ \partial_\alpha \delbar_{\bbar} 
\varphi $ where $\eta_{\alpha \bbar}$ 
is a flat 
background metric,  
(\ref{constrG}) becomes 
the following equation for $\varphi$: 
\begin{equation} 
\det \left( \begin{array}{cc} 1+\partial_1 \delbar_1 \varphi &  \partial_1 \delbar_2 
\varphi \\ \partial_2 \delbar_1 \varphi & -1+\partial_2 \delbar_2 
\varphi \end{array} \right) =1 
\label{Pleq} 
\end{equation} 
in the notation of~\cite{OV}. This equation and~(\ref{feO}) can be derived from 
the action 
\begin{equation} 
\int \partial \varphi \delbar \varphi +\frac{1}{3!} \varphi \delbar \varphi 
\wedge \partial \delbar \varphi +\int \sqrt{G}G^{\alpha \bbar}
\partial_\alpha \Omega \partial_{\bbar} \Omega ,
\label{Pleb} 
\end{equation} 
where the first term is the Plebanski action (also with the notation of~\cite{OV})
and the second term is such that~(\ref{feO}) is the field equation 
obtained by varying $\Omega$ and using the constraint~(\ref{constrG}). The
action~(\ref{Pleb}) can be thought of as the dual action. 
 
We have thus established the following result. 
We started with the theory of hermitian gravity with torsion in four dimensions 
defined by the 
action~(\ref{KMHact}), 
the field equations of which implied that the curvature with torsion was 
anti-self-dual and 
with holonomy   $SU(2)$ (for signature (4,0)) or $SL(2,R)$ (for signature (2,2)). 
We then dualised this to obtain anti-self-dual Riemannian gravity coupled to
a harmonic scalar $\Omega$, with no 
torsion and the action~(\ref{Pleb}). Thus in four dimensions, a theory 
with torsion is related by a conformal rescaling to a theory without torsion.
This is in agreement with the results of~\cite{CHS,Tod}. We 
emphasize  that this duality (unlike the dualities considered 
e.g. in~\cite{RV,GHR,LR,O}) does not 
require any Killing vectors.

        The generalisation to other (even) dimensions is straightforward. Consider 
the action 
\begin{equation} 
S= \int d^D x \left[ \sqrt{|g|} 
-\frac{1}{4}  \left\{ 
\Lambda^{\alpha \bbar} \left( g_{\alpha \bbar}- 
\partial_{\alpha} \kbar_{\bbar} -\delbar_{\bbar} k_{\alpha} \right) 
+c.c. \right\} 
 \right] .
\label{S-Lagr} 
\end{equation} 
Eliminating $\Lambda^{\alpha  
\bbar}$ from~(\ref{S-Lagr}), we recover the action~(\ref{KMHact}). Alternatively, 
integrating out $k_{\alpha}$, $\kbar_{\abar}$ gives the constraints~(\ref{constr}). The 
solution to~(\ref{constr}) in $D=2n$ dimensions is 
\begin{equation} 
\Lambda^{\alpha \bbar} = L^{\alpha \bbar} 
\label{sol2n} 
\end{equation} 
where 
\begin{equation} 
L^{\alpha \bbar} \equiv \epsilon^{\alpha \gamma_1 \ldots 
\gamma_{n-1} \bbar \overline{\delta}_{1} \ldots \overline{\delta}_{n-1}} 
\partial_{\gamma_1} \dbar_{\overline{\delta}_{1}} K_{\gamma_2 \ldots 
\gamma_{n-1} \overline{\delta}_{2} \ldots \overline{\delta}_{n-1}} 
\label{K2n} 
\end{equation} 
is the \lq field strength' of an $(n-2,n-2)$ form $K$. The action then takes the form 
\begin{equation} 
S= \int d^D x \left[ \sqrt{ |g|} 
-\frac{1}{4}  \left\{ 
L^{\alpha \bbar} g_{\alpha \bbar} 
+c.c. \right\} 
 \right] 
\label{Ssolved} 
\end{equation} 
with $L$ given by~(\ref{K2n}). The field equation for $g_{\alpha \bbar}$ 
is 
\begin{equation} 
\sqrt{|g|} g^{\alpha \bbar} = L^{\alpha \bbar}  .          \label{feD} 
\end{equation} 
Taking determinants in~(\ref{feD}), we find 
\begin{equation} 
\sqrt{|g|} = \left| \det L^{\alpha \bbar} \right|^{\frac{1}{n-2}} 
\label{detD} 
\end{equation} 
Contracting~(\ref{feD}) with $g_{\alpha \bbar}$  yields 
\begin{equation} 
L^{\alpha \bbar} g_{\alpha \bbar} = n   \left| \det L^{\alpha \bbar}  
\right|^{\frac{1}{n-2}} . 
\label{Lambdag} 
\end{equation} 
It is easily checked that the solution of the field equation~(\ref{feD}) 
is of the form 
\begin{equation} 
g^{\alpha \bbar} = \mu \left| 
\det L^{\alpha \bbar} \right|^\nu L^{\alpha \bbar} 
\end{equation} 
where $\mu$ and $\nu$ are constants given by 
\begin{equation} 
\mu=1 \ \ , \ \ \nu = -\frac{1}{n-2} . 
\end{equation} 
Substituting~(\ref{detD}) and~(\ref{Lambdag}) into~(\ref{Ssolved}) gives 
the dual action 
\begin{equation} 
S= -\frac{1}{2} (n-2) \int d^D x \left|  \det L^{\alpha \bbar} 
\right|^{\frac{1}{n-2}}    
\label{Sdual2n} 
\end{equation} 
for the field strength $L$ of the $D-4$ form potential $K$. Again, we have chosen the Lagrange multiplier 
to be a tensor density so that the Lagrange multiplier term in the action is coordinate invariant.

	We now reinstate the fixed background $\hat{g}_{\alpha \bbar}$, which will be taken to be of 
the form $\hat{g}_{\alpha \bbar} =  
\delbar_{\bbar} \hat{k}_{\alpha} 
+\partial_{\alpha} 
\hat{\kbar}_{\bbar}+i\Upsilon_{\alpha \bbar}$ and includes the coupling to Yang-Mills fields, through  
$\Upsilon_{\alpha \bbar}$. 
As a result of
(\ref{F2&U}),
this background metric satisfies
\begin{equation} 
\ghat_{\alpha [ \bbar ,\gambar ] \delta} -
\ghat_{\delta [ \bbar ,\gambar ] \alpha } 
=-4 F_{\alpha [ 
\bbar }F_{\gambar ] \delta} .
\label{modif} 
\end{equation} 
 Consider the action 
\begin{equation} 
S = \int d^D x \left[ \sqrt{ |g|}  
-\frac{1}{4} \left\{ \Lambda^{\alpha \bbar}  
\left( g_{\alpha \bbar} -\hat{g}_{\alpha \bbar}  
-\partial_{\alpha} \kbar_{\bbar} -\delbar_{\bbar} k_{\alpha} \right) + c.c. \right\} \right] . 
\label{constact} 
\end{equation} 
Eliminating $\Lambda^{\alpha \bbar}$ from~(\ref{constact}), we recover the  
action~(\ref{KMHact}) (after shifting the potentials $k \to k - \hat k$). 
The vectors 
are Lagrange multipliers for the 
constraints~(\ref{constr}), which can be solved locally in terms of a $D-4$ form 
$K$ as 
in~(\ref{sol2n}).  On integrating out the vectors, the action takes the form 
\begin{equation} 
S= \int d^D x \left[ \sqrt{ |g|} 
-\frac{1}{4}  \left\{ L^{\alpha \bbar}  
\left( g_{\alpha \bbar} -\ghat_{\alpha \bbar} \right) +c.c. \right\} \right] 
\label{heract4} 
\end{equation} 
where $L^{\alpha \bbar}$ is given in~(\ref{K2n}).  
Using the field equation for $g_{\alpha \bbar}$, taking the determinant and the  
trace and substituting 
back into the action~(\ref{heract4}), we find the dual action in the form 
\begin{equation} 
S= \frac{1}{4} \int d^D x \left[ L^{\alpha \bbar}\ghat_{\alpha \bbar} +c.c.
-2(n-2) 
\left| \det L^{\alpha \bbar} \right|^{\frac{1}{n-2}}  \right] .
\label{dual2nYM} 
\end{equation} 
($n\neq 2$). In the absence of Yang-Mills fields,  
$\Upsilon_{\alpha \bbar}=0$, then the term 
$L^{ij}\ghat_{ij}$ in~(\ref{dual2nYM}) 
vanishes after  integration by parts, as a result of  
the form~(\ref{K2n}) of $L^{\alpha \bbar}$  
and the fact that the background metric $\ghat_{\alpha \bbar}$ 
satisfies~(\ref{2ndconstr}).

If the Yang-Mills fields do not vanish,
then using the form~(\ref{K2n}) of $L^{\alpha \bbar}$, integrating by 
parts and using~(\ref{modif}), we obtain 
\begin{equation} 
S= - \frac{1}{2} (n-2) \int d^D x \left|  \det L^{\alpha \bbar} 
\right|^{\frac{1}{n-2}}   - \frac{1}{2}  \int K \wedge tr (F\wedge F) 
\label{Sdualt5t5} 
\end{equation} 
with an interesting coupling of the $D-4$ form potential to $ F\wedge F$.  
 
For $n=2$, 
integrating out the metric in~(\ref{heract4}) gives
the constraint
\begin{equation}
\det L^{\alpha \bbar} =1    \label{again}
\end{equation}
for signature (2,2), or $\det L^{\alpha \bbar} =-1$ for signature (4,0), while 
the action reduces to the term  $\int K \wedge tr (F\wedge F)$.
The constraint~(\ref{again}) can be imposed via a Lagrange multiplier $\Lambda$, so that the action becomes
\begin{equation} 
S= - \frac{1}{2}  \int   K \wedge tr (F\wedge F)   -\frac{1}{2}
\int d^D x \Lambda \left( \det
L^{\alpha \bbar} -1 \right)  .
\label{St5lamb}
\end{equation} 
Integrating out the scalar $\Lambda$ yields the constraint~(\ref{again}), so that one
recovers the action $\int K\wedge (F\wedge F )$ subject to this constraint. Instead we keep the
Lagrange multiplier; using~(\ref{F2&U}), (\ref{K2n}) and integrating by parts, we find the 
following field equation for $K$
\begin{equation}
\partial \delbar \left( i\Upsilon -\Lambda \det (L^{\alpha \bbar})
L^{-1} \right) = 0 , 
\end{equation}
where $L^{-1}$ is the 2-form $(L^{-1})_{\alpha \bar \beta}
dz^\alpha \wedge d{\bar z}^{\bar \beta}$. 
This implies that
\begin{equation}
\Lambda (\det L )L^{-1} = iJ' 
\label{feKk'}
\end{equation}
where
\begin{equation}
J'
=i( \partial \kbar ' +\delbar k' ) +\Upsilon
\end{equation}
for some (1,0) form potential $k'$,
so that $J'$ is the 2-form corresponding to a  metric $g'_{\alpha \bar \beta}$ defining some
dual (2,1) sigma-model.
 Although~(\ref{feKk'}) is not algebraic in $K$, one can solve for
$\Lambda$ as a functional of $K$, $k'$ and $F$; taking determinants
in~(\ref{feKk'}), we find
\begin{equation}
\Lambda = \pm  \sqrt{ \frac{\det \left(g' \right)}{\det L}} ,
\end{equation}
which can be
substituted back in~(\ref{St5lamb}).

It is useful to 
define a dual metric $\tilde{g}_{\alpha \bbar}$ (for $n \ne 1$)  by  
\begin{equation} 
\tilde{g}_{\alpha \bbar} \equiv \left| \det L^{\alpha \bbar}  
\right|^{\frac{1}{n-1}}  \left( L^{-1} \right)_{\alpha \bbar} , 
\label{newmet} 
\end{equation} 
so that  
\begin{equation} 
L^{ij}= \sqrt { 
\det {\tilde g_{ij}} 
} \tilde g^{ij} .
\end{equation} 
Then  
the dual geometry is given in terms of the fundamental two-form 
\begin{equation} 
\tilde{J} = -i \tilde g_{\alpha \bbar} dz^\alpha \wedge d\bar z ^{\bar \beta} 
\end{equation} 
by 
\begin{equation} 
\tilde{J} = i * \partial \delbar K 
\label{newkah} 
\end{equation} 
where the Hodge star operation is defined with respect to  
the metric~(\ref{newmet}). The dual action~(\ref{Sdualt5t5}) (for $n\ne 2$) can also be  
expressed in terms of the dual metric~(\ref{newmet}) and we find 
\begin{equation} 
S= - \frac{1}{2}(n-2) \int d^D x \left|  \det \tilde g _{\alpha \bbar} 
\right|^{\frac{n-1}{n-2}}   -\frac{1}{2} \int * \tilde{J} \wedge \Upsilon 
\label{Sdualt5t5etr} 
\end{equation} 
using eqs.~(\ref{F2&U}) and~(\ref{newkah}). The constraint~(\ref{newkah}) 
defines a class of hermitian geometries (without torsion) in which the metric is given in terms of a $D-4$ 
form potential instead of the scalar potential of K\"{a}hler geometry.   
The action~(\ref{Sdualt5t5etr}) gives a field equation for such metrics which arises naturally from the 
dualisation of (2,1) geometry. 

For $n=2$, the K\"{a}hler form $J_G=i \partial \delbar K $ corresponding to
the K\"{a}hler metric $G_{\alpha \bar \beta}$ defined in~(\ref{defG}) is dual to $\tilde J$, $J_G=*\tilde
J$, so that the action   becomes
\begin{equation} 
S=  -\frac{1}{2} \int * \tilde{J} \wedge \Upsilon =
-\frac{1}{2} \int  J_G \wedge \Upsilon
\label{Sdualt5t5etdadaadadr} 
\end{equation} 
subject to the constraint~(\ref{newkah});
this is the Donaldson action~\cite{Don} for self-dual Yang-Mills in a self-dual geometry.

	The field equation for the Yang-Mills fields derived from the action~(\ref{Sdualt5t5etr}) is
\begin{equation}
\tilde{J}^{ij}F_{ij} =0 .
\label{genUY}
\end{equation}
Thus $F$ satisfies the Uhlenbeck-Yau
equation with respect to the complex structure $\tilde{J}_{ij}$.
This can be derived, for example, by 
transforming with a complex gauge 
transformation with parameter $h$ taking values in the complexification $G_c$ of the
gauge group $G$~\cite{H1},
\begin{equation}
F=h^{-1}fh
\end{equation}
where
\begin{equation}
f=da +a^{2} = \delbar a
\end{equation}
is the field strength of a holomorphic connection
given by the  (1,0)  form $a=U^{-1}\partial
U$.
Then using $trF^2 = tr f^2$ and varying with respect to the prepotential $U$ gives
\begin{equation}
\partial \delbar   K\wedge f  +\delbar K \wedge \partial f
+\delbar K \wedge [a,f] =0
\label{kf0}
\end{equation}
which gives~(\ref{genUY})
on using~(\ref{newkah}) and the Bianchi identity.

 The field equations for the dual metric $\tilde{g}_{ij}$ and its implications for the dual geometry
will be discussed elsewhere.
Note that the dualisation procedure carried out in the foregoing 
can also be applied to the actions of section~\ref{sec:pol} with an auxiliary metric; the results 
are 
equivalent.

\section{Conclusion}

        Summarizing, the (2,1) sigma-model or string give rise to a theory of 
gravity coupled to a two-form gauge field, a dilaton and Yang-Mills gauge 
fields 
in $D=2n$ dimensions. The field equations imply that the curvature with torsion 
is self-dual in four dimensions, or has $SU(n)$ holonomy in $2n$ dimensions. 
The 
system is described by the Born-Infeld type action~(\ref{KMHact}), where 
$g_{\alpha 
\bbar}$ is given in terms of $k_\alpha$ by~(\ref{21geom}). This action can be 
simplified using an auxiliary metric, and the forms~(\ref{Hform}) 
and~(\ref{Hform4}) are 
classically equivalent to~(\ref{KMHact'}) and~(\ref{KMH4}) respectively. The 
four-dimensional action~(\ref{Hform4}) is classically  
invariant under diffeomorphisms preserving the volume element constructed from the  
auxiliary metric and under the 
generalised Weyl transformation~(\ref{Weyl}).  
It would be interesting to compare this symmetry   to the 
infinite dimensional current algebra~\cite{LMNS} of the Donaldson 
action for self-dual Yang-Mills~\cite{Don,NS}.  
 
The action~(\ref{KMHact'}) can be dualised, with no isometries 
being required. The dual theory  in four dimensions is self-dual gravity {\em without} torsion 
coupled to a scalar. This 
recovers the remarkable equivalence between self-dual hermitian 
geometries with torsion and self-dual gravity without torsion.  In higher 
dimensions, dualising gives an aparently new generalisation of 
K\"{a}hler geometry, in which the
metric $\tilde g_{ij}$ is hermitian  
and is determined by the 
$(n-2,n-2)$ form potential $K$ (which can be thought of as  analogous to the scalar
potential of  K\"{a}hler geometry) via~(\ref{K2n})
and the dynamics is described 
by 
the action~(\ref{Sdualt5t5etr}).  The couplng to Yang-Mills is via the term $K\wedge F\wedge F$
and gives rise to the Uhlenbeck-Yau-type field equation~(\ref{genUY}).

\vspace{1.5cm} 
 
{\bf Acknowledgements} 
\\ 
\\ 
        We would like to thank Emil Martinec for helpful comments. M.\ A.\ 
would also like to thank Jos\'{e} Figueroa-O'Farrill for 
useful discussions. The work of M.\ A.\ is supported in part by the Overseas 
Research Scheme and in part by Queen Mary and Westfield College, London. 
C.\ M.\ H.\  is supported by an EPSRC Senior Fellowship.

\end{document}